# Localized dynamics arising from multiple flat bands in a decorated photonic Lieb lattice

**HAISSAM HANAFI,[1,*] PHILIP MENZ,[1] ALLAN MCWILLIAM,[1,2] JÖRG IMBROCK,[1] AND CORNELIA DENZ[1]**

[1] *Institute of Applied Physics, University of Münster, Corrensstr. 2, 48149 Münster, Germany*
[2] *Department of Physics, University of Strathclyde, 107 Rottenrow East, Glasgow G4 0NG, United Kingdom*
[*] *haissam.hanafi@uni-muenster.de*

**Abstract:** Photonic lattices have emerged as a promising approach to localize light in space, for example, through topologically protected edge states and Aharonov-Bohm caging. They are of particular importance in the study of flat band systems via the associated nondiffracting compact localized states. However, such states typically appear static, thus not allowing adaptive or evolutionary features of light localization. Here we report on the first experimental realization of an oscillating compact localized state (OCLS). We observe an oscillatory intensity beating during propagation in a two-dimensional photonic decorated Lieb lattice. The photonic system is realized by direct femtosecond laser writing and hosts multiple flat bands at different eigenenergies in its band structure. Our results open new avenues for evolution dynamics in the up to now static phenomenon of light localization in periodic waveguide structures.

Wave localization is a highly active area of research involving various disciplines of physics with its origin being the seminal work of Anderson in 1957, where he describes the *'Absence of Diffusion in Certain Random Lattices'* [1]. Intriguingly, it has been shown that periodic lattices with particular geometries also allow complete localization of the wave function within a finite region [2–5]. Such systems are distinguished by completely flat or dispersionless bands in their band structure. A flat band implies the degeneracy of a macroscopic number of momentum eigenstates at the flat band energy. In other words, the entire flat band has a vanishing wave group velocity and describes single-particle states with an infinite effective mass, which renders them effectively motionless. As a result of the quenched kinetic energy, flat bands are considered to be ideal playgrounds to study strongly correlated phenomena. For example, they have been linked to the fractional quantum Hall effect [6], and to high-temperature superconductivity [7, 8]. Moreover, the degeneracy of the flat band eigenstates allows their combination to obtain states that are strictly localized, meaning that they have zero amplitude in real space due to destructive interference at all but a few lattice sites. These states are known as compact localized states (CLSs).

Distinct CLSs have been experimentally demonstrated in a number of different systems [9], including metamaterials [10], Bose-Einstein condensates [11], polariton condensates [12], acoustic lattices [13], and photonic lattices [3–5, 14, 15]. The latter consist of evanescently coupled waveguides and yield spatial nondiffracting flat band states. Light propagation in photonic lattices is governed by a Schrödinger-like paraxial wave equation, where the propagation constant $\beta$ takes the role of the energy, and the band structure can be interpreted as a spatial diffraction relation [16]. The *temporal* evolution of an electronic wavefunction in an atomic potential is therefore mapped onto the *spatial* evolution along the propagation direction in the photonic lattice. CLSs, as the signature of flat bands, manifest themselves through light localization, i.e., they propagate in the photonic lattice without diffraction. Most photonic realizations of CLSs have focused on simple lattices hosting a single flat band, for example, in the Lieb [3, 4], or kagome lattices [5]. In these systems, it is possible to linearly combine CLSs from the same flat band in order to obtain dispersionless states of nearly arbitrary form. This mechanism has been proposed for distortion-free image transmission [17, 18]. More recently, lattices hosting multiple flat bands have been proposed in 1D [19, 20], and 2D [21–23], and the bands' singularity and nonsingularity have been studied [15]. The presence of multiple flat bands in the band structure of a single lattice system opens up a new research field, as it allows a study of wave dynamics of superimposed CLSs originating from flat bands at different eigenenergies. This leads to the question of what to expect from such a superposition of spectrally-separated nondiffracting states, and whether it is possible to observe their behavior experimentally.

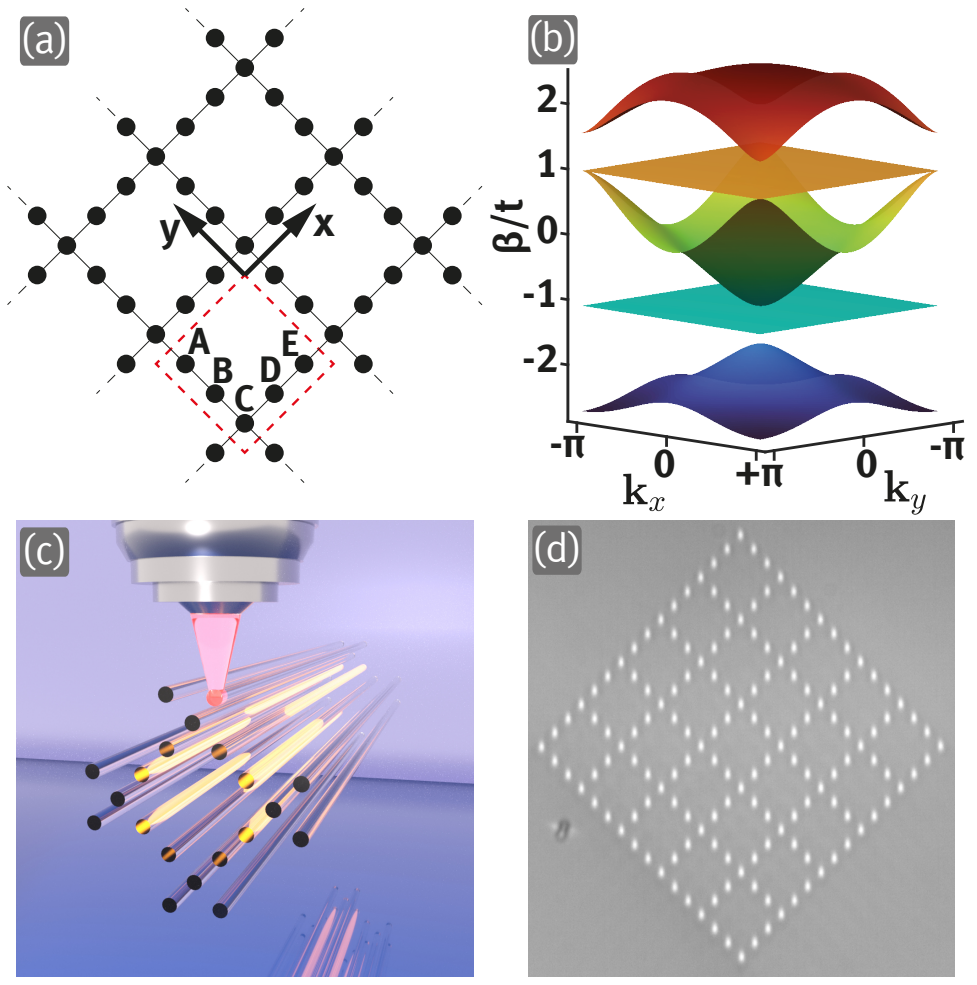

Fig. 1. (a) Schematic of the Lieb-5 lattice, with the unit cell marked by the dotted red lines and each individual lattice site labeled from A to E. (b) band structure calculated using the tight-binding model. (c) Femtosecond direct laser writing technique. (d) micrograph of the fabricated lattice.

In this letter, we tackle this question and report, to the best of our knowledge, on the first experimental observation of an oscillating compact localized state (OCLS). The oscillation manifests itself as a longitudinal intensity beating during propagation in the two-dimensional photonic lattice. The OCLS is constructed from flat band states at different eigenenergies, with the energy difference of the flat bands defining the spatial oscillation frequency. This case is in stark contrast to the previously investigated fundamental CLS, as well as to superpositions of CLSs originating from the same flat band, since they inherently display static diffractionless propagation in a photonic lattice. The oscillation of the OCLS allows observing for the first time unusual evolution dynamics compared to the otherwise static phenomenon of light localization.

Among the lattices hosting multiple flat bands, perhaps the simplest one that only involves short-range hopping and does not need staggered fluxes or periodic driving consists of an extension of the Lieb lattice [24–26]. The Lieb lattice is the 2D counterpart of the 3D perovskite or edge-centered cubic lattice. Intriguingly, some high-temperature superconductors exhibit a Lieb lattice structure in the copper oxide planes of cuprates, and the flat band has been hypothesized to be the origin of their high critical temperature [8, 27]. The Lieb lattice can be interpreted as a decoration of a standard square lattice adding one lattice site between every corner. It is possible to extend this decoration further by adding two lattice sites between the existing ones of the square lattice. Having a total of five lattice sites per unit cell, this structure is denoted as the Lieb-5 lattice and is illustrated in Fig. 1(a). We label the respective lattice sites from A to E. The double decoration leads to a shift in the flat band energy compared to the simple

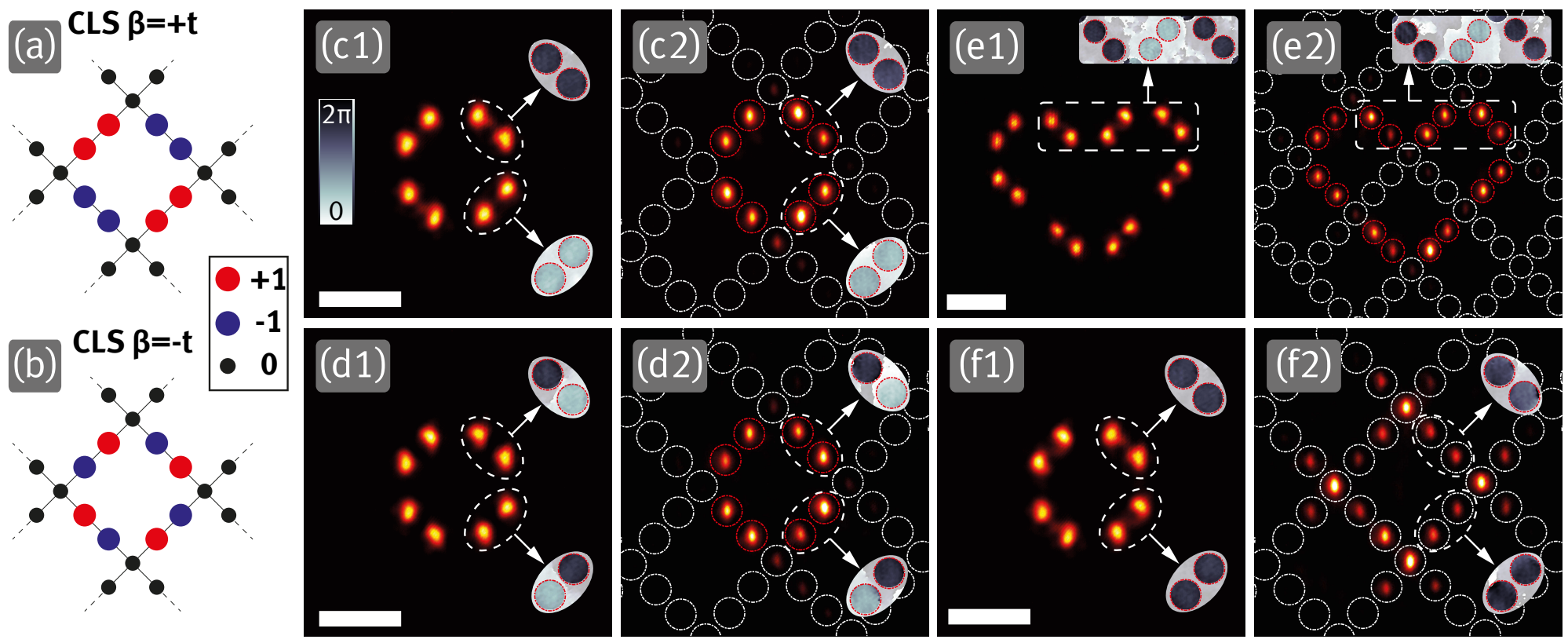

Fig. 2. Observation of diffractionless propagation of CLSs in the photonic Lieb-5 lattice. The scalebars correspond to 50 µm. (a) CLS of the flat band at $\beta = +t$. (b) CLS of the flat band at $\beta = -t$. (c1) Experimentally measured intensities (the insets showing the corresponding phases) of light field corresponding to the CLS at $\beta = +t$. (c2) same as (c1), but after a propagation of 2 cm in the lattice. (d1)-(d2) same as (c1)-(c2), but for the CLS at $\beta = -t$. (e1)-(e2) same as (c1)-(c2), but for the combination of three translated CLSs at $\beta = +t$. (f1)-(f2) same as (c1)-(c2), but for the diffracting state having equal phase at all lattice sites as shown in the insets.

Lieb lattice [28].

By calculating the band structure following a tight-binding model with only nearest neighbor coupling Fig. 1(b), two flat bands are obtained. These are located at $-t$ and $+t$ respectively, where $t$ is the nearest neighbor coupling strength. This intriguing situation raises the possibility of investigating particle-hole excitations in the form of exotic heavy excitons [21].

Each of these flat bands implies the existence of strictly localized eigenstates at their respective propagation constants. The CLSs of the Lieb-5 lattice are decorations of the four-site ring CLSs of the conventional Lieb lattice. They fulfill the condition of destructive interference at the minority sublattice sites (sites labeled C in Fig. 1(a)) by having the same amplitudes, and crucially a $\pi$-phase difference in the nearest neighboring majority sublattice sites. The CLSs take the form of octagonal rings of lattice sites which are pairwise either in-phase or out-of-phase as shown in Fig. 2(a) and (b), respectively [24].

In the following, we experimentally demonstrate the diffractionless nature of the CLSs in a photonic waveguide-based realization of the Lieb-5 lattice. We achieve this through the femtosecond direct laser writing technique and fabricate an array of evanescently coupled waveguides in fused silica. This flexible and versatile technique permits a realization of arbitrary lattice geometries and has been employed to realize various flat band lattices [3, 4, 15, 29]. We fabricate a Lieb-5 lattice with a flat edge termination composed of 156 waveguides. The front facet of the sample is shown in the microscope image in Fig. 1(d). The length of each waveguide is 2 cm in the direction of propagation $z$, with each inscribed waveguide spaced at a separation distance of $d = 22$ µm with respect to its nearest neighbors. This separation distance is optimal as it minimizes the next-nearest neighbor coupling which is detrimental to the flatness of the bands, however, still allows for sufficient nearest neighbor coupling during propagation in the lattice. We excite the flat band states using structured laser light with a wavelength of $\lambda = 532$ nm coupled into the corresponding waveguides. The appropriate amplitude and phase modulation corresponding to the CLSs is achieved using a spatial light modulator. After propagation through the lattice, we image the output light states at the back facet of the lattice and recover their respective phase information by interfering them with a plane wave [30]. In this way, we verify that the flat band states propagate diffractionless in the Lieb-5 lattice. When launching the CLS of the flat band at $\beta = +t$ into the lattice (see Fig. 2(c1)), the light remains localized at the initially excited lattice sites during propagation (see Fig. 2(c2)). To evaluate the degree of localization quantitatively, we evaluate the intensity of the propagated light at each lattice site. We find that 95% of the total intensity remains localized in the initially excited waveguides. As shown in the insets representing the phase at the lattice sites, highlighted by the white ellipses, the input phase relation leading to destructive interference at the corner sites is conserved. The results for the CLS of the second flat band at $\beta = -t$ are shown in Fig. 2(d). The input field displayed in (d1) propagates in the lattice without diffracting (d2) as the phase difference of $\pi$ is preserved. In this case, 93% of the total intensity remains localized at the excited lattice sites. To validate the nondiffracting nature, we implement a light field with the same intensity but with a homogeneous phase at all lattice sites (Fig. 2(f1)). In this case, diffracting bands as well as the flat bands are clearly excited. As a result, after propagation in the lattice, the light field is no longer localized due to strong diffraction caused by coupling to initially unexcited waveguides (Fig. 2(f2)).

The linear superposition of CLSs from the same flat band allows obtaining diffractionless states that can take arbitrary shapes, i.e., of any desired image, with the smallest building block being the fundamental CLS. We exemplary demonstrate the power of this approach using a state with the shape of a heart (Fig. 2(e1)), which is formed by the superposition of three spatially-shifted CLSs from the flat band at $\beta = +t$. The distortion-free image transmission after propagation in the lattice is shown in (Fig. 2(e2)). 94% of the total intensity remains localized in the initially excited lattice sites. The inset shows that the phase structure leading to destructive interference at all C lattice sites is well preserved.

In a final visionary approach, we study the dynamics of superimposed CLSs originating from two different flat bands (Fig. 3). As the flat bands are located at distinct propagation constants $\beta$, we expect the superimposed states to form oscillating compact localized states (OCLS) with an angular frequency of $\Delta\beta = 2t$ in the propagation direction $z$. In our experimental realization, we image the back facet of the lattice and therefore the output light field after a fixed propagation distance of $z = 2\,\mathrm{cm}$. To fully capture the spatial dynamics of the OCLS, we virtually propagate the input field and image the output for different inputs [31]. As the propagation constants of the CLSs are $\beta = +t$ and $\beta = -t$, they will acquire a phase factor of $e^{\pm i t z}$ during propagation in the lattice, resulting in an intensity beating in propagation direction $\left|e^{itz} + e^{-itz}\right|^2 \sim \cos^2$. Therefore, as sketched in Fig. 3 (a), we vary the phases of the two CLSs of the input light field to achieve a virtual propagation. We then image the output light fields, which reveal the oscillating behavior, as shown in Fig. 3 (b) (bottom row). Due to the real propagation of $z = 2\,\mathrm{cm}$ in the photonic lattice, additional phase factors of $e^{\pm it\cdot 2\,\mathrm{cm}}$ are acquired by the two CLSs from which we estimate a difference between the propagation constants of the flat bands of $\Delta\beta = 2t \approx 1\,\mathrm{cm}^{-1}$. To highlight the oscillations that occur, Fig. 3 (c) shows the measured intensity in the waveguides labeled D (light gray) and E (dark gray) in Fig. 3 (b) (bottom row) during the virtual propagation. The dot markers show the experimental data, while the solid line shows the expected $\cos^2$ function. The observed oscillations demonstrate that diffractionless propagation of CLSs in photonic lattices is not necessarily a static phenomenon but can be adaptively tailored to exhibit dynamics by controlling the angular frequency via the band structure.

In conclusion, we have experimentally established a photonic Lieb-5 lattice using the femtosecond direct laser writing technique. We demonstrated diffractionless propagation of the CLSs associated with the two flat bands in the spatial band structure of the lattice. We additionally showed that superpositions of CLSs belonging to the same flat band lead to diffractionless states of any desired shape. Importantly, we revealed the existence of OCLS originating from flat bands at different eigenenergies that propagate in the lattice without diffraction but exhibit a longitudinal oscillatory beating in intensity. Our results significantly advance the experimental realization of compact localization in multi flat band systems and bring an exciting new dynamic into the picture. They may provide inspiration for extending studies of flat band physics to the unexplored scenario of multiple flat bands. Diffractionless propagation in photonic lattices lends itself towards applications in distortion-free image transmission which is significantly enriched by the possibility of obtaining patterns that exhibit spatial dynamics using the revealed OCLS.

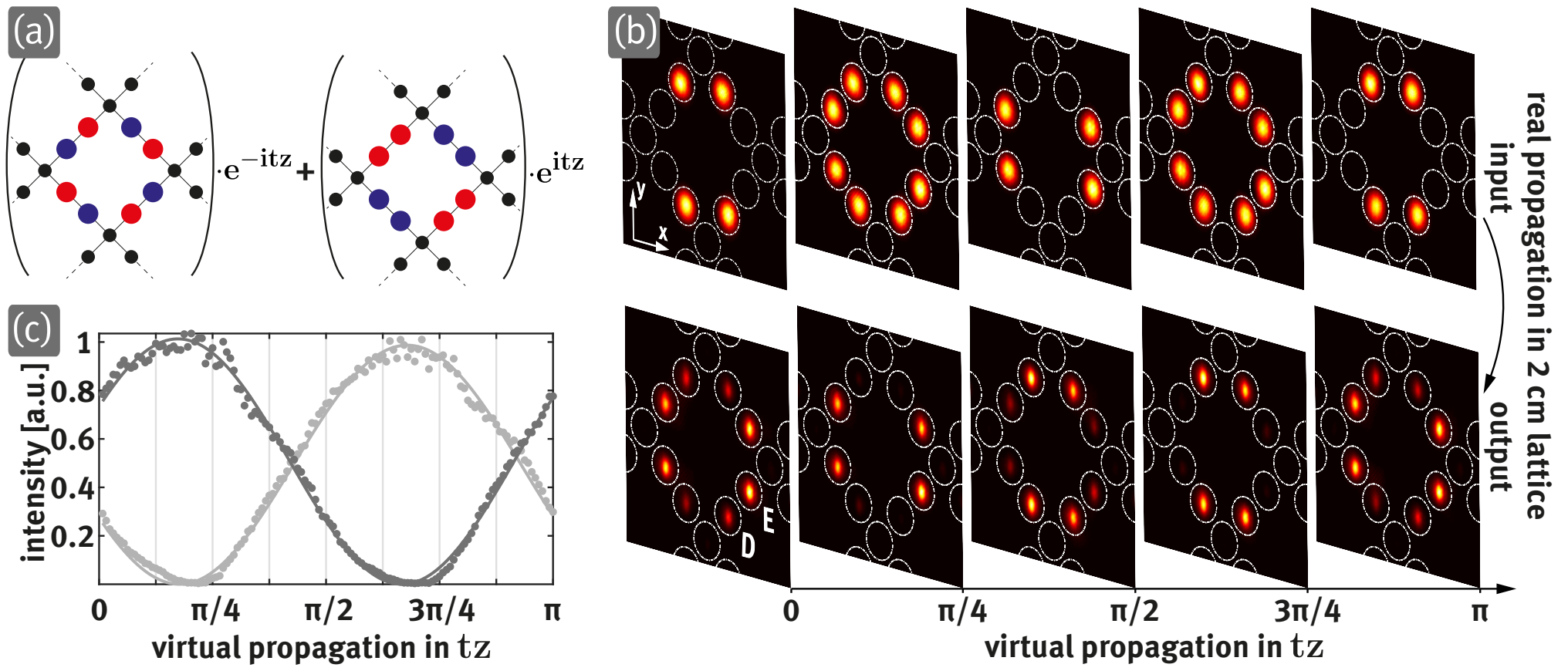

Fig. 3. Experimental results of the propagation dynamics of the OCLS. (a) The OCLS is formed by the linear superposition of CLSs from the two flat bands at $\beta = \pm t$. By varying the phase between the two CLSs by a factor of $e^{\pm itz}$ respectively, different input states are obtained. This corresponds to a virtual propagation in the lattice. Transverse intensity measurements of different input states are shown in (b) (top row). The corresponding output intensities, after a real propagation in the photonic lattice, are shown in (b) (bottom row). (c) displays the intensity guided by the waveguides labeled with D (light gray) and E (dark gray) in (a). The dot markers show the experimental data, and the solid line shows the expected $\cos^2$ function.